\documentstyle[11pt,paspconf,psfig]{article}

\begin{document}

\title{Early results of the ESO VLT}
\author{S.Cristiani}
\affil{Space Telescope European Coordinating Facility, European
Southern Observatory, Karl Schwarzschild Strasse 2, D-85748 Garching,
Germany }

\begin{abstract}
The results of the FORS and ISAAC Science Verification 
of the FORS and ISAAC instruments at the VLT ANTU/UT1 are described.
The following observations have been carried out:
1) the Cluster Deep Field  MS1008.1-1224
2) the Antlia dwarf spheroidal Galaxy
3) multiple Object Spectroscopy of Lyman break galaxies in
the AXAF and Hubble Deep Field South
4) ISAAC IR Spectroscopy of a gravitationally magnified galaxy
at z=2.72.
The data have been made public for the ESO community and, in the case
of HDF-S, worldwide. 
\end{abstract}


\section{Introduction}
The general goals that ESO intended to reach with Science Verification 
(SV) Observations were manifold (Giacconi et al. 1999): 
{\it i)} to experiment with scientific observing runs, including the end-to-end
VLT dataflow (Silva and Quinn, 1997),
{\it ii)} to produce data of scientific quality, involving as early as
possible the ESO community in their analysis,
{\it iii)} to foster an early scientific return from the VLT,
{\it iv)} to obtain feedback about the telescope performances and operational 
procedures.

To this end an SV team has been formed at ESO, under the leadership of
A.Renzini.
A first block of SV observations has been carried out in August 1998
with the VLT test camera. Results have been published in Vol. 343
of the  A\&A Journal.
A second block of SV observations was planned to be executed in
January and February 1999 and in particular with the two
instruments for optical/IR imaging and spectroscopy: FORS1 (Nicklas et 
al. 1997)
and ISAAC (Moorwood 1997).
The FORS-1/ISAAC SV Team includes the following scientists: J. Alves,
S. Cristiani, R. Hook, R. Ibata, M. Kissler-Patig, P. M{\o}ller,
M. Nonino, B. Pirenne, R. Rengelink, A, Renzini, P. Rosati, D. Silva,
E. Tolstoy, and A. Wicenec.
The actual schedule of the SV observations was more complex than
originally planned and benefitted from observations executed by the 
FORS1 and ISAAC Commissioning Teams.

FORS1 and ISAAC are rather complex instruments, with many observing
modes. The goal of the SV has been to cover the main modes, selecting
scientific programmes of outstanding interest.  
The following observations are briefly described here:
\begin{enumerate}
\item{The Cluster Deep Field MS1008.1-1224}
\item{The Antlia dwarf spheroidal Galaxy}
\item{Multiple Object Spectroscopy (MOS) of Lyman break galaxies in
the AXAF and Hubble Deep Field South}
\item{ISAAC IR Spectroscopy of the gravitationally magnified galaxy
at z=2.72 MS1512cB58.}
\end{enumerate}

\section{The FORS-ISAAC Cluster Deep Field MS1008.1-1224}

\begin{figure}
\psfig{figure=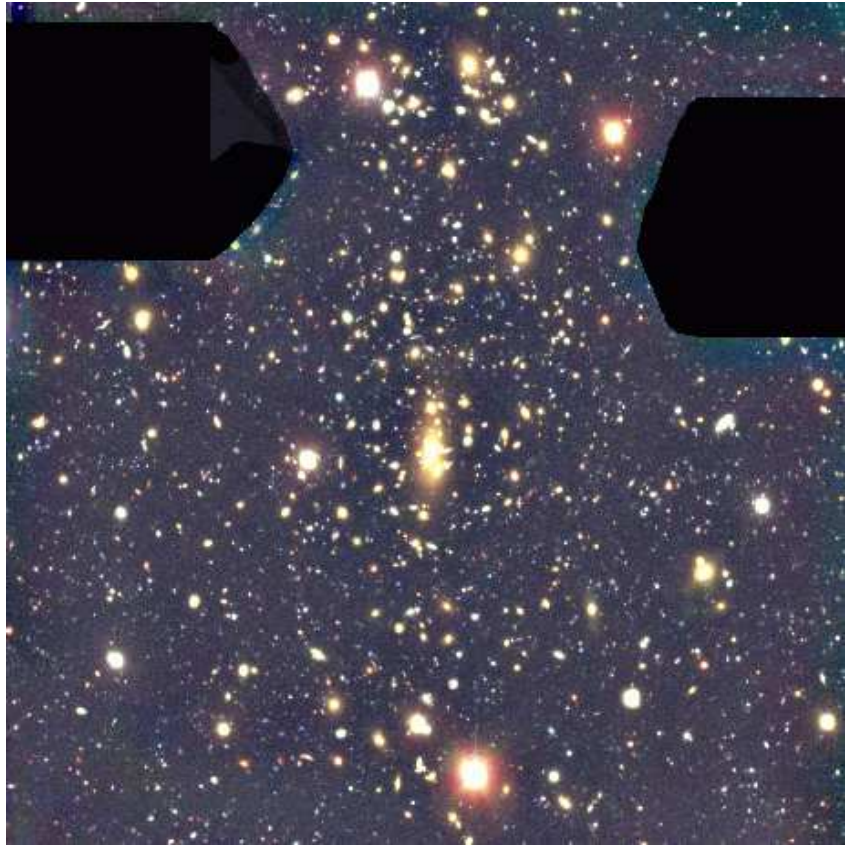,height=9.7cm}
\caption{Combined color image of the Cluster MS1008.1-1224}
\label{fig-clust}
\end{figure}

The criteria inspiring the selection of the cluster were:
existence of published data indicating a large mass/velocity
dispersion at redshift=$0.3-0.5$, existence of gravitational arcs, 
optimal visibility throughout most of the night in January-March
(i.e. RA=8-11hr).
The selection narrowed on the X-ray selected cluster MS1008.1-1224 (RA
= 10 10 32.2, DEC =--12 39 55. ep.2000) from the Einstein Medium
Sensitivity Survey (EMSS, Gioia and Luppino, 1994) at z=0.30, also
part of the CNOC Survey (Carlberg et al. 1996).

BVRIJK observations were carried out with total integration times
between 4000 and 5400 s in the optical and about 1h in the IR.
The K-band observations were obtained under exceptional seeing
conditions and produced de-jittered coadded images of $0.4''$ PSF.

These data will be used for:
{\it a)} a detailed study of the cluster mass distribution from
gravitational lensing shear maps, magnification bias and strong
lensing features,
{\it b)}{ a study of the cluster galaxy population down to $\sim 4$ mag below
L$^*$, using color-mag diagrams}, 
{\it c)} {a search for highly magnified distant galaxies}, 
{\it d)} {obtaining photometric redshifts of all galaxies in the field}, 
{\it e)} {identifying interesting stars in the field.}

\section{The Antlia Dwarf Spheroidal Galaxy}

\begin{figure}[t]
\psfig{figure=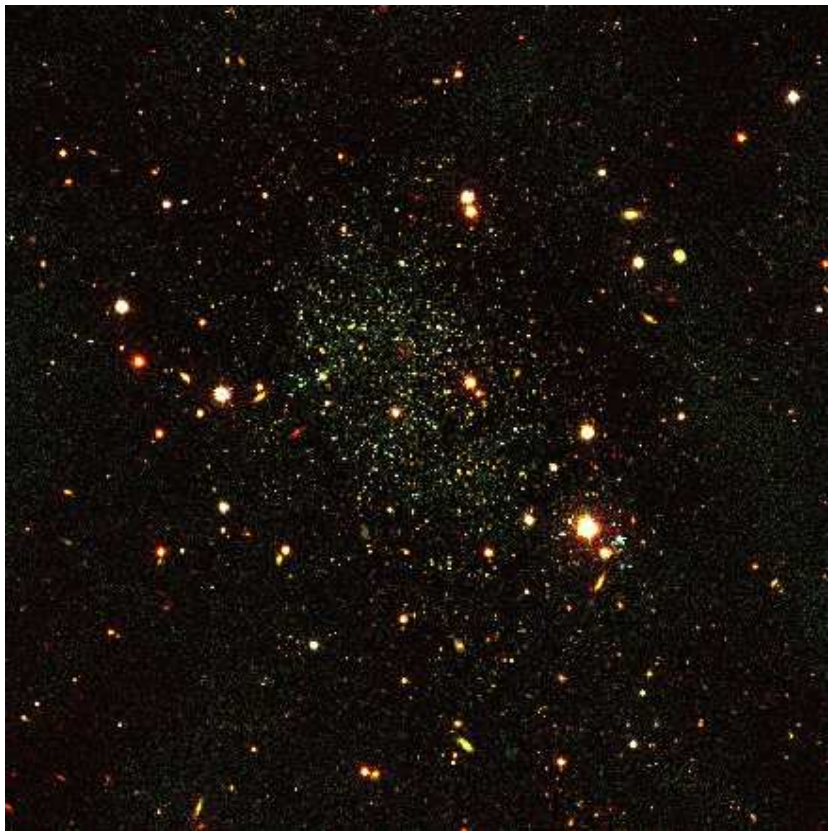,height=9.7cm}
\caption{Combined color image of the Dwarf Galaxy Antlia}
\label{fig-Antlia}
\end{figure}

Antlia is a small low surface brightness dwarf spheroidal type galaxy
(RA= 10 04 04, DEC=--27 19 49, ep. 2000, l=263, b=22)
first noted in an HI survey of southern hemisphere galaxies by Fouqu\'e
et al. (1990), which was discovered in 1997 to be an outlying member
of the Local Group. 
This galaxy was selected for the SV because:
published data indicate a distance close enough to allow the detection 
of resolved stars well down the Red Giant Branch, there is a
controversy about the existence and importance of a young stellar
population and reddening in Antlia and the visibility is optimal in 
January-March.

Images have been obtained with FORS in B ($4 \times 600 s$), V ($4
\times 600 s$) and I ($18 \times 300 s$), all with a PSF close to
$0.5''$. The combined color image is shown in Fig.~\ref{fig-Antlia}.

These data will be used for:
{\it i)} a detailed study of the Colour-Magnitude diagrams to determine the
properties of the resolved stellar population in  this small nearby,
relatively isolated galaxy,
{\it ii)} determining if there is a young population from the presence or
absence of a Main Sequence,
{\it iii)} comparing the properties of a more distant dwarf spheroidal
galaxy with those we see around our Galaxy,
{\it iv)} a study of the reddening properties of this system with
colour-colour diagrams,
{\it v)} confirming the distance to this system.

\section{Multiple Object Spectroscopy of Lyman break galaxies in
the AXAF and Hubble Deep Field South}

\begin{figure}[t]
\psfig{figure=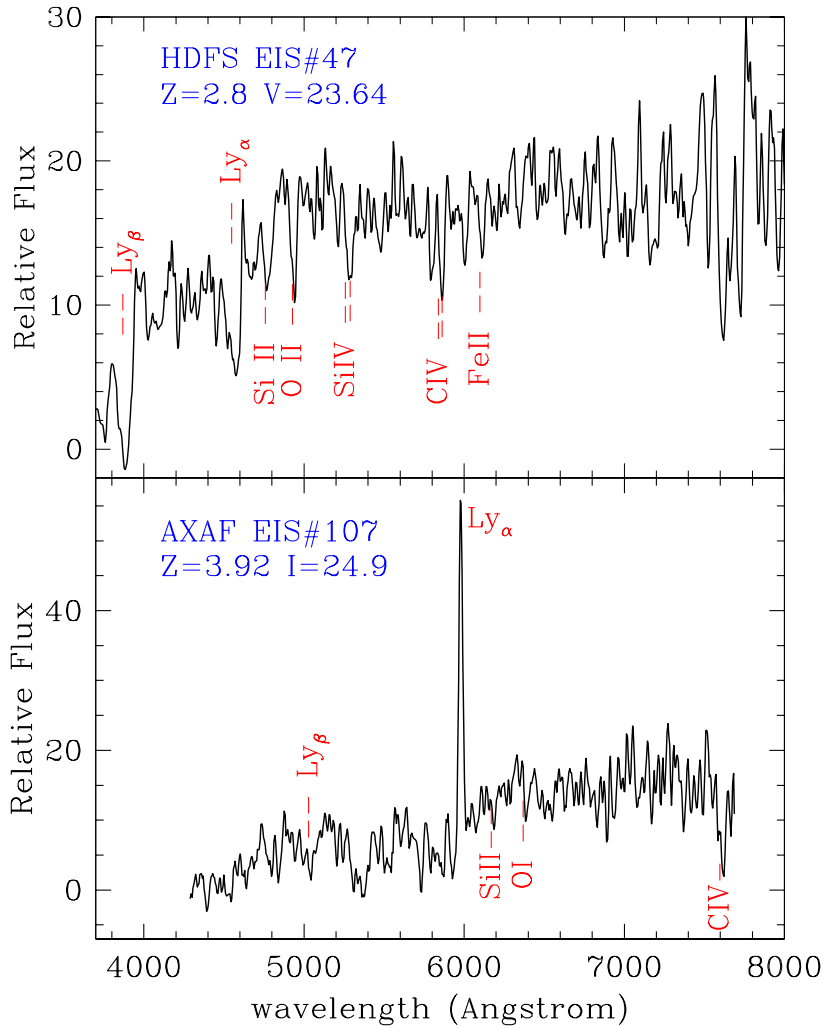,height=13cm}
\caption{Spectra of two high-z galaxies in the AXAF and Hubble Deep
Field South}
\label{fig-MOS}
\end{figure}

The study of Lyman-break galaxies was the main spectroscopic
programme of the FORS1 Science Verification.
The science goals are:
{\it i)} to push the FORS1 MOS to the limits and optimize its
observational strategy and data reduction procedures for the specific
case of faint galaxies,
{\it ii)} to provide pilot observations for future studies aiming at
characterizing the abundance of high-redshift galaxies by luminosity,
size, morphology, star formation rate and clustering,
{\it iii)} to check and refine the photometric selection criteria for
high-z galaxies,
{\it iv)} help to tune future wide-angle optical surveys to maximize
the yield of Lyman-break galaxies,
{\it v)} to provide galaxies in suitable redshift ranges for future
spectroscopic observations with ISAAC.

The AXAF field (Giacconi et al., 1998) was selected for SV
observations, given the
availability of deep multi-color imaging obtained with SUSI-2 and SOFI
and derived catalogues produced by the EIS project. 
Data of the same
type were also publicly available for a field including the WFPC-2
pointing in the Hubble Deep Field South. Targets in this latter field
were provided to the FORS1 commissioning team and were observed by
them in December 1998. 
Targets were extracted from the lists of U and B dropouts given in
Table 5 and Table 6 of Rengelink, R. et al. (1998) for the AXAF field 
and in Table 9 of da Costa, L. et al. (1998) for the HDF-S.
When no suitable candidate was available for the allowed range of
positions of a given slit, a random object in the field was chosen.
Spectra of the objects in both the AXAF field
and HDF-S have been reduced and analyzed in the same way
within the MIDAS package.
A total of 8 galaxies with redshifts between 2.8 and 4 were
identified.
The spectra of two of them are shown in Fig.~\ref{fig-MOS}.
A full table and details of the reduction are given at the URL
{\tt http://http.hq.eso.org/science/ut1sv/MOS\_index.html}. 

\section{Low-resolution spectroscopy in the H band of MS1512-cB58}

\begin{figure}
\psfig{figure=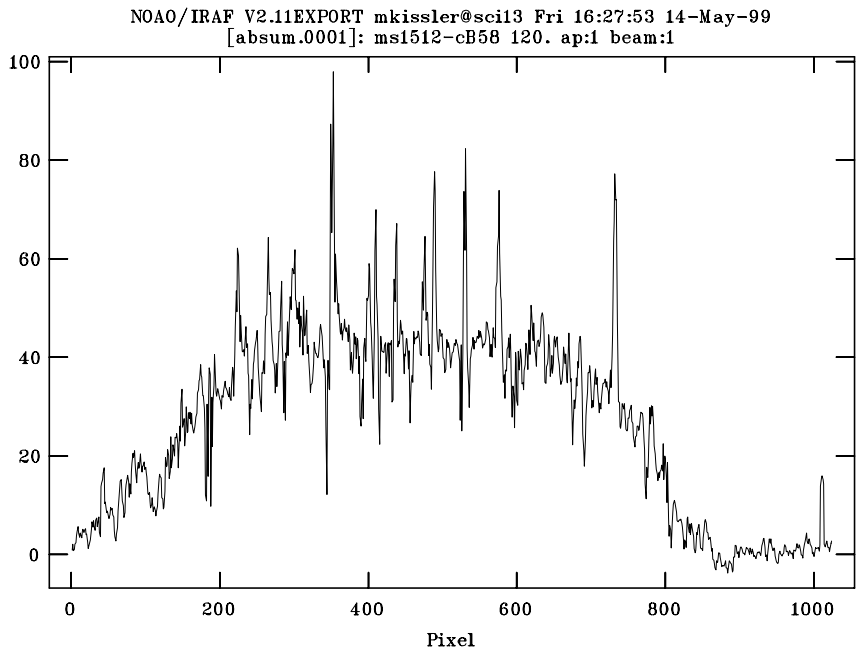,height=8cm}
\caption{Uncalibrated ISAAC spectrum of the galaxy MS1512-cB58}
\label{fig-cb58}
\end{figure}

The gravitationally lensed galaxy 
MS1512-cB58 was serendipitously discovered by Yee et al. (1996),
in the course of the CNOC cluster survey. It lies in the
field of the z=0.36 cluster MS1512+36, and its redshift was derived
from a dozen strong absorption lines in the rest frame UV (Ellingson
et al. 1996).  Most interestingly, this galaxy is
gravitationally amplified by a factor of about 30. Its
near-IR magnitudes are J=19.12, H=18.42, K'=17.83, i.e. the object
has exceptionally bright apparent magnitudes for its redshift.
Therefore, it offers a
unique opportunity to get a high S/N infrared spectrum of a high
redshift galaxy, with important lines such as $H_\alpha$ (at the edge of
the K band) and $H_\gamma$ (within the H band) being uncontaminated by
atmospheric emission/absorption. The detection of such lines would
allow a more reliable determination of the star formation rates,
compared to other indicators such as [O II] and the UV
continuum. More importantly, the profile of such lines - if present -
may provide hints on the mass of this high-redshift galaxy.  ISAAC is
the first instrument that offers this opportunity, and the unusually
large amplification factor made it an attractive target for the VLT,
in spite of the large zenithal distance at which it can be observed
from Paranal.
The data were obtained in the night of March 24, 1999. ISAAC was
operated in the short-wavelength, low-resolution mode.  The SH filter
was used (covering 1.4 to 1.82 micron), with the $1''$ slit. The
resolution in this configuration is around 500.  The pixel scale is
$0.147''$/pixel. A preliminary data reduction was carried out with a
final exposure of 5040 sec effective exposure time. The result is
shown in Fig.~\ref{fig-cb58}.
\section{Data Access}
The data have been made public for the ESO community and, in the case
of HDF-S, worldwide. Full details and data request forms can be found
at the URL: {\tt http://http.hq.eso.org/science/ut1sv/}.


\begin{references}
\reference Carlberg, R.G., Yee, H.K.C., Ellingson, E.; Abraham, R.,
Gravel, P., Morris, S., Pritchet, C. J., 1996, ApJ, 462, 32
\reference da Costa, L. et al. 1998, A\&A submitted, astro-ph/9812105
\reference Ellingson, E., Yee, H.K.C., Bechtold, J., Elston, R.,
1996, ApJ 466, L71
\reference Fouqu\'e, P., Durand, N., Bottinelli, L.,
Gouguenheim, L., Paturel, G., 1990, A\&AS 86, 473
\reference Giacconi R et al., Proceedings of "Highlights of X-ray
Astronomy", in honour of J.Truemper's 65th birthday Garching, Germany,
17-19 June 1998 Ed. B.Aschenbach
\reference Giacconi R., Gilmozzi R., Leibundgut B., Renzini A.,
Spyromilio J., Tarenghi M., 1999, A\&A 343, L1
\reference Gioia I., Luppino, G.A., 1994, ApJS 94, 583
\reference Moorwood A., 1997, SPIE 2871, 1146
\reference Nicklas, H., Seifert, W., Boehnhardt, H., Kiesewetter-Koebinger, S.,
Rupprecht, G.,  1997, SPIE 2871, 1222
\reference Rengelink, R. et al. 1998, A\&A submitted, astro-ph/9812190 
\reference Silva D., Quinn, P., 1997, The Messenger 90, 12
\reference Yee, H.K.C., Ellingson, E., Bechtold, J., Carlberg, R.G.,
Cuillandre, J.-C., 1996, AJ, 111, 1783
\end{references}
\end{document}